\documentclass[aps,showpacs,twocolumn,amsmath,amssymb,pra]{revtex4-1} 
 
\usepackage{graphicx} 
\usepackage{bm}

\begin{document} 
 
\title{Chaos-assisted, broadband trapping of light in optical resonators} 
 
\author{C. Liu$^1$, A. Di Falco$^2$, D. Molinari$^{1,3}$, Y. Khan$^1$, B. S. Ooi$^1$, T. F. Krauss$^2$ and A. Fratalocchi$^1$} 

\email{andrea.fratalocchi@kaust.edu.sa} 
\homepage{www.primalight.org}

\affiliation{
$^1$PRIMALIGHT, Faculty of Electrical Engineering; Applied Mathematics and Computational Science, King Abdullah University of Science and Technology (KAUST), Thuwal 23955-6900, Saudi Arabia\\
$^2$School of Physics and Astronomy, University of St. Andrews, North Haugh, St. Andrews KY16 9SS, UK  \\
$^3$Dept. of Astronomy, Bologna University, via Ranzani 1, I-40127, Bologna, Italy.
}

\date{\today} 
 
\begin{abstract} 
Chaos is a phenomenon that occurs in many aspects of contemporary science. In classical dynamics, chaos is defined as a hypersensitivity to initial conditions. The presence of chaos is often unwanted, as it introduces unpredictability, which makes it difficult to predict or explain experimental results. Conversely, we demonstrate here how chaos can be used to enhance the ability of an optical resonator to store energy. We combine analytic theory with ab-initio simulations and experiments in photonic crystal resonators to show that a chaotic resonator can store six times more energy than its classical counterpart of the same volume. We explain the observed increase with the equipartition of energy among all degrees of freedom of the chaotic resonator, i.e. the cavity modes, which is evident from the convergence of their lifetime towards a single value. A compelling illustration of the theory is provided by demonstrating enhanced absorption in deformed polystyrene microspheres.
\end{abstract} 
 
 
\maketitle 

\subsection*{Introduction}
The enhanced interaction between light and matter in optical cavity resonators is a subject of a great interest that impacts on many areas of condensed matter physics such as cavity electrodynamics \cite{mabuchi02:_cavit_quant_elect}, quantum and nonlinear optics, but also on more applied aspects such as optical signal processing \cite{bourzac12:_photon_chips_made_easier,kengo12:_ultral_power_all_optic_ram_based_nanoc,leo10:_tempor_cavit_solit_in_one} and resonantly enhanced optical absorption \cite{atwater10:_plasm_for_improv_photov_devic}. 
All these applications are enabled by highly optimized optical resonators that can efficiently trap electromagnetic energy in narrow frequency bands. In conventional cavities, there is a simple trade-off between bandwidth and the enhancement of trapped energy; the higher the enhancement, the narrower the bandwidth. A great challenge in the field is therefore to develop a new generation of cavities that are able to break this fixed relationship and to store more energy in a given bandwidth window than conventional cavities would allow.\\ 
The maximum power that can be transferred into a conventional resonator depends on the coupling coefficient and loss of each given mode and tends to vary across the mode spectrum, especially when broadband operation ($\Delta$ $\lambda$ $\approx$ 100's of nm) is considered. In addition, classical 2D and 3D geometries tend to accommodate modes with very different lifetimes in the same spectral region, a good example being the widely used photonic crystal "L3" type cavity that features modes of very different Q-factor closely spaced in frequency \cite{chalcraft:241117}. Therefore, the use of classical resonators for broadband energy storage is limited. Here, we overcome this intrinsic limitation by exploiting specific shape deformations that support chaotic trajectories for light rays. Surprisingly, we note that in a chaotic cavity, the lifetimes of all the modes tend towards a common value, thus improving the transfer of energy into the cavity and increasing the energy-storage capability of the cavity. Such chaotic resonators \cite{campillo95:_optic_proces_in_microc} have been well exploited in the field of laser devices
\cite{PhysRevLett.94.233901,noeckel97:_ray_and_wave_chaos_in,PhysRevLett.90.063901,Gmachl05061998,
PhysRevLett.99.224101,PhysRevLett.101.084101,PhysRevLett.102.044101,PhysRevLett.78.4737}. We also note that the mode spectrum of deformed microsphere resonators has been studied recently with remarkable changes in the Q-factors being observed \cite{murugan12:_optic_microd_reson_by_flatt_micros}. However, despite this large body of literature, nothing is known about the capacity of such resonators to store and collect light energy over a broad spectral range.\\ 
The increased energy storage capacity of a chaotic resonator compared to a classical one can be intuitively explained by adopting a ray optics approach and considering that a suitable shape deformation is accompanied by the breaking of symmetry in the structure.  As a consequence, the deformed resonator cannot support any cyclic motion of light, thus the trajectory of light rays changes from regular to random, statistically resulting in a larger lifetime of the photons in the cavity \cite{campillo95:_optic_proces_in_microc}. In order to clarify this result, we start our analysis from a symmetric (classical) resonator and observe its capacity to trap energy as it is deformed. We define a deformation parameter, $\alpha$, which we will use in the following description as a handle to deform any given geometry, with $\alpha=0$ indicating the original, undeformed structure and $\alpha>0$ indicating a proportionally deformed geometry. For example, a circle would be described by $\alpha=0$, and a deformed circle by $\alpha>0$, with $\alpha$ describing the degree of deformation. This parameter is generic and can be used to describe the deformation of any type of resonator, e.g. square or disk in 2D or cube or sphere in 3D. In general, the larger $\alpha$, the larger the degree of chaos, until a saturation value is reached. The parameter $\alpha$ is defined in Eq. (\ref{curve}) below.\\
We begin our analysis with a single \emph{ab-initio} numerical experiment, which demonstrates a six-fold increase of the energy stored inside a chaotic resonator with respect to a classical counterpart of the same volume. 
Such an analytic treatment allows us to study the lifetimes of the electromagnetic modes excited in the cavity and to examine the coherent buildup of energy inside the system. We find that energy gets uniformly distributed among the spectral degrees of freedom and that the cavity is able to store the same amount of energy for a given wavelength interval.
We support our theoretical results with a set of experiments both in 2D and 3D geometries. 
Two-dimensional resonators were fabricated in planar photonic crystals (PhCs)  and analyzed by pump and probe transmission measurements. We observed and characterized the resulting modal lifetimes for varying degrees of chaos. 
While the PhC environment provides fine control over the shape of the resonator, hence its chaotic behavior, it is difficult to measure the energy stored inside the cavity directly.
For this reason, we also studied a three-dimensional arrangement of polystyrene spheres, each suitably modified  to exhibit chaotic behavior. Despite the high transparency of polystyrene, absorption measurements showed a broadband absorption enhancement of $\approx 10\%$ across its entire absorption bandwidth ($\approx 450$nm), due to the larger energy trapped in the chaotically deformed resonator. It is worthwhile to stress that the 
only reason for the increased amount of energy trapped in the resonator is the onset of chaos caused by the deformation of the spheres, with the amount of material being kept strictly constant.
\begin{figure}
\centering
\includegraphics[width=7.5cm]{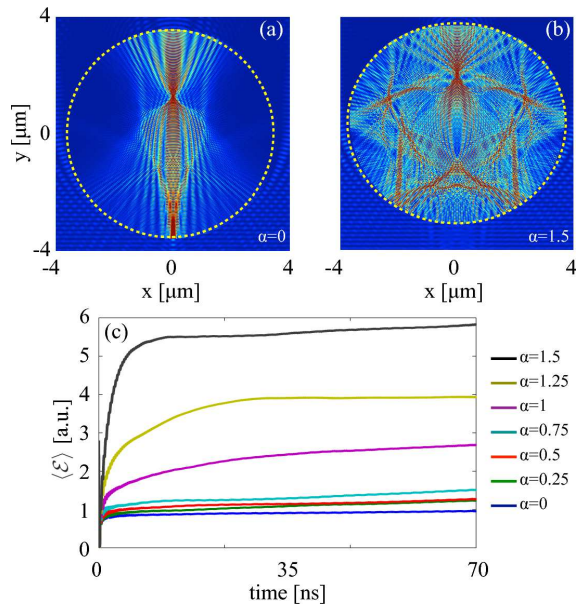}
\caption{
\label{sample} (a) Snapshot of the electromagnetic energy density $\mathcal{H}$ distribution 
after $t=45 fs$ in a resonator (dashed line) with $V=30\mu m^2$ for (a) $\alpha=0$ and (b) $\alpha=1.5$; (c) time 
evolution of the average electromagnetic energy $\langle\mathcal{E}\rangle$ for different $\alpha$.}
\end{figure}
\subsection*{A single ab-initio experiment: broadband light confinement and energy equipartition} 
The theory is more easily developed in 2D starting from a circular resonator.   
Our setup consists of a silicon dielectric resonator with air holes , whose shape is defined
 by the following function in polar coordinates $(\rho_c,\theta_c)$:
\begin{align}
\label{curve}
&\rho_c=\sqrt{\frac{A}{\pi}-\frac{\alpha^2}{2}}+\alpha\cos\theta_c, &0\le\theta_c\le 2\pi,
\end{align} 
with $A$ the resonator area and $\alpha\ge 0$ the single parameter that controls the resonator shape. Equation (\ref{curve}) belongs to the family of analytic curves investigated by 
 Robnik: for $\alpha>0$,
Eq. (\ref{curve}) supports chaos in the trajectory of light rays, which randomly bounce 
inside the resonator \cite{0305-4470-16-17-014}. From a physical perspective, the shape defined 
by Eq. (\ref{curve}) is equivalent to an asymmetric deformation of a disk, and can be realized experimentally with conventional nanofabrication tools.\\ 
In our simulations, we fixed the resonator area to $A=30\mu m^2$ and numerically calculated the electromagnetic energy $\mathcal{E}$ stored inside the resonator for varying values of $\alpha$. Although the value of $\alpha$ in Eq. (\ref{curve}) is not bound to an upper value, we note that the system reaches saturation for a maximum degree of chaos described by $\alpha=1.5$ (see Materials). We therefore restrict our numerical analysis to $\alpha\in[0,1.5]$. The calculation of the electromagnetic energy $\mathcal{E}(t)=\int_\mathcal D\mathcal{H}d\mathbf{\rho}$ is performed by a numerical integration of the energy density $\mathcal{H}(\rho;t)=\frac{1}{2}(\mathbf{E}\cdot\mathbf{D}+\mathbf{H}\cdot\mathbf{B})$ in the volume $\mathcal{D}$ encompassed by the resonator and defined by $\rho\le\rho_c$ and 
$0\le\theta\le 2\pi$. We simulated the input from a supercontinuum source in the wavelength range between $\lambda=300nm$ and $\lambda=1300nm$, which simulates a broadband source such 
as sunlight. Figure \ref{sample}a-b shows a time snapshot of the spatial distribution 
of $\mathcal{H}$ after $t=45$ fs, and illustrates how the geometry of the resonator changes with $\alpha$ (Fig. \ref{sample}a-b, dashed line). As seen, a small deformation in the shape (at a constant volume) already yields a radically different behavior in the distribution of light energy inside the resonator.
Figure \ref{sample}c, conversely, displays the time-averaged energy $\langle\mathcal{E}\rangle=\frac{1}{t}\int_0^t dt'\mathcal{E}(t')$ 
evolution with increasing parameter $\alpha$.
Quite remarkably, the introduction of chaos into the motion of light is accompanied by a dramatic change of the energy stored inside the resonator, whose steady state regime 
---when the insertion of energy balances radiation losses--- shows an $\approx$ 6-fold increase already 
for $\alpha=1.5$ (Fig. \ref{sample}b). This energy accumulation grows with the 
deformation, and monotonically rises as $\alpha$ is increased from $\alpha=0$ to $\alpha=1.5$.\\ 
\begin{figure*}
\centering
\includegraphics[width=14cm]{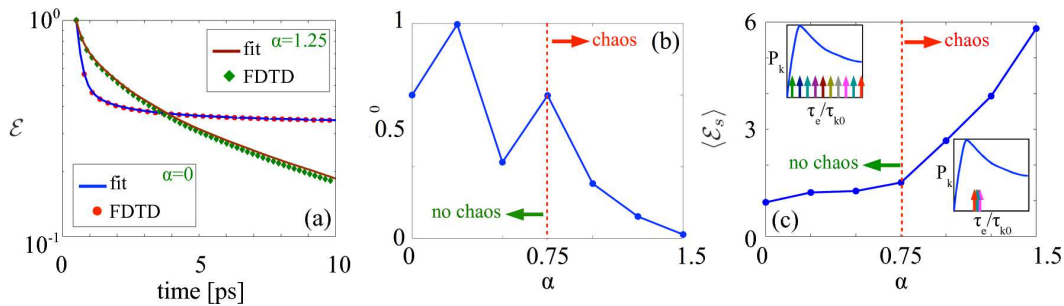}
\caption{
\label{com} (a) Log-plot of the energy $\mathcal{E}$ relaxation dynamics for $\alpha=0$ and $\alpha=1.25$: FDTD results (markers) and Prony exponential fits (solid lines); (b) normalized distribution of the difference between the maximum and minimum decay constants $\Delta(\alpha)/\Delta(0)\equiv\Delta/\Delta_0$ versus $\alpha$; (c) steady state energy distribution $\langle\mathcal{E}_s\rangle=\langle\mathcal{E}\rangle(t=45 fs)$ for varying values of $\alpha$. The insets in (c) sketch the effects of the convergence of the decay constants $\tau_k$ on the power $\mathcal{P}_k$ transferred into the structure, whose behavior versus $\tau_e/\tau_{k0}$ is plotted as a solid line; in the non-chaotic situation (left) the distribution of $\tau_e/\tau_{k0}$ (left, colored arrows) is very broad and only a few frequencies are able to efficiently transfer power; in the strongly chaotic case, conversely, a condensed distribution of $\tau_e/\tau_{k0}$ yields the same contribution for all wavelengths and a much larger number of frequencies can contribute to storing energy in the structure.
}
\end{figure*}
We use Time Dependent Coupled Mode Theory (TDCMT) \cite{HausBook} to develop a simple and reliable model for the light-resonator interaction.
The system can be modeled as a side coupled resonator, whose dynamical equations can be easily solved to get the energy $\mathcal{E}_k$ and the power $\mathcal{P}_k$ stored into the the $k-$th mode (see Materials):
\begin{align}
\label{tdcmt}
&\mathcal{E}_k=\frac{\tau_{k}^2}{\tau_e}\big(1-e^{-\frac{t}{\tau_{k}}}\big)^2, &
\mathcal{P}_k=\frac{2\frac{\tau_e}{\tau_{k0}}}{\big(1+\frac{\tau_e}{\tau_{k0}}\big)^2}|S|^2,
\end{align}
with $|S|$ the input source power, $\frac{1}{\tau_k}=\frac{1}{\tau_{k0}}+\frac{1}{\tau_e}$ the mode decay rate, with $1/\tau_{k0}$ the intrinsic cavity 
decay rate of the k-th mode and $1/\tau_e$ the escape rate 
due to coupling with the environment. 
The power $\mathcal{P}_k$ strongly depends on the ratio between the radiation and the 
coupling loss through the parameter $\tau_{k0}/\tau_e$, achieving the 
maximum value of $\mathcal{P}_k=0.5|S|^2$ when $\tau_e=\tau_{k0}$.
Outside this matching condition, the power coupled into the structure decreases very fast. 
To study how the dynamics of the decay rates can be affected by chaos, we began by calculating the evolution of the decay rates $\tau_{k0}$ for different $\alpha$ in the resonator defined by Eq. (\ref{curve}). In a series of FDTD simulations,  
we first excited the resonator with a source and then monitored 
the energy evolution $\mathcal{E}(t)$ when the source was switched off. The decay constants $\tau_{k0}$ are extracted from 
the time energy evolution $\mathcal{E}(t)=\sum_k|A_k|^2 e^{-2t/\tau_{k0}}$ by applying the Prony method \cite{tijhuis87:_elect_inver_profil}. Fig. \ref{com}a shows two typical examples of numerical simulations (markers) and fit (continuous line) for two different values of $\alpha$. In general, modes of different frequency exhibit different  decay rates $1/\tau_{k0}$. 
However, when strong chaos is generated in the structure, the distribution of $\tau_{k0}$ converges towards
a frequency independent delta function. In fact, calculations which start from initial 
 conditions that already belong to the chaotic region of phase space will always be distributed according 
 to the same probability distribution, which defines the so called \emph{natural invariant measures} 
 of the system \cite{TOCS}. As a result, when the entire phase-space is dominated by chaos, 
 we observe the same evolution of the decay rates for all possible initial conditions towards a frequency-independent decay rate $\tau_{k0}=\tau_0$. A convenient way of highlighting this dynamics is to plot the difference between the maximum and the minimum decay constant: 
 $\Delta(\alpha)=\max(\tau_{k0})-\min(\tau_{k0})$, for different values of $\alpha$ (Fig. \ref{com}b). 
 We clearly observe a transition scenario: below $\alpha^*$, the dynamics shows simply an 
 oscillation of $\Delta(\alpha)$ around the same average value, while above the chaos 
 threshold $\alpha\ge\alpha^*$, a clear convergence of $\Delta\rightarrow 0$ is observed. The value $\alpha^*$ depends 
 on the specific geometry of the chaotic resonator, and can be assessed by calculating the relative 
 area of the system phase space that encompasses chaos (see Materials).
The effect of this convergence towards a single lifetime of all the modes on the energy collected by the resonator can be readily 
 evaluated from Eq. (\ref{tdcmt}). The total power transferred into the structure, in particular, 
 then becomes frequency independent $\mathcal{P}_k=\mathcal{P}_0=2|S|^2\big(1+\tau_e/\tau_{0}\big)^{-2}\tau_e/\tau_{0}$, 
 and every mode contributes to the same extent to storing energy inside the resonator. 
 When many modes are present in the resonator, their large number results in 
 a coherent buildup process that leads to a significant accumulation of energy (Fig. \ref{sample}b and Fig. \ref{com}c).
In the non-chaotic case, conversely, much fewer modes are able to efficiently transfer energy 
into the resonator due to the mismatch between $\tau_{k0}$ and $\tau_e$ (Fig. \ref{com}c, inset) and 
the system can store relatively less energy (Fig. \ref{sample}b and Fig. \ref{com}d).\\  
\begin{figure}
\centering
\includegraphics[width=8cm]{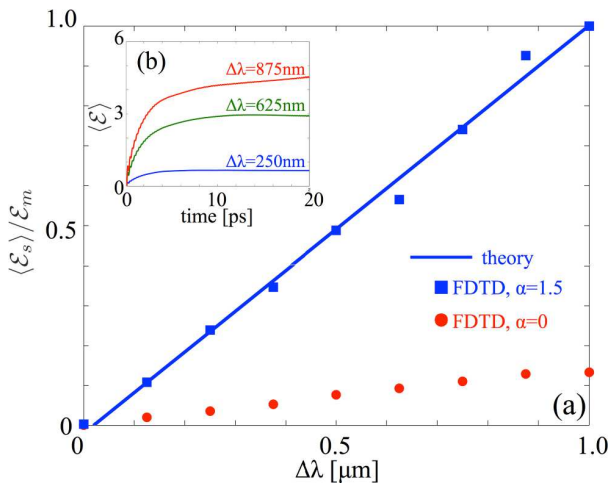}
\caption{Results for a variable bandwidth source: (a) FDTD calculated average energy $\langle\mathcal{E}\rangle$ versus time for $\alpha=1.5$ and different normalized bandwidth $\Delta\lambda$ symmetrically centered at $800$nm; (b) FDTD computed steady state average energy $\langle\mathcal{E}_s\rangle$ versus bandwidth $\Delta\lambda$ for $\alpha=0$ (circle markers) and $\alpha=1.5$ (square markers). In (a) the solid line indicates the behavior predicted by Eq. (\ref{eth}). The energy $\langle\mathcal{E}_s\rangle$ is normalized to the maximum value $\mathcal{E}_m$ attained for $\alpha=1.5$ and $\Delta\lambda=1\mu$m.  
\label{eqp} 
}
\end{figure}
\begin{figure*}
\centering
\includegraphics[width=14cm]{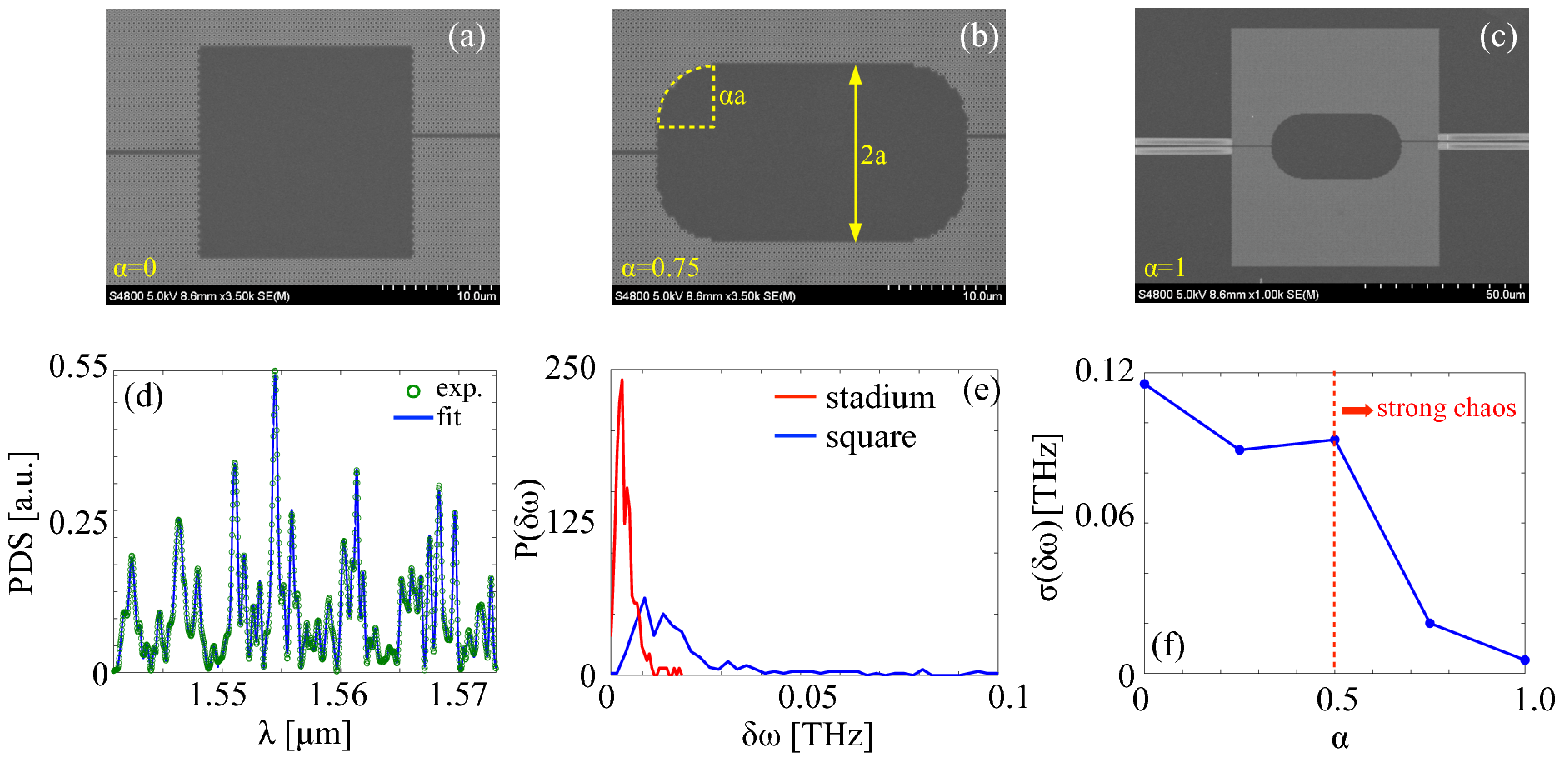}
\caption{Summary of 2D experimental results: (a)-(c) SEM micrographs of the sample geometry for (a) $\alpha=0$, (b) $\alpha=0.75$ and (c) $\alpha=1$; (d) experimental spectrum (circle markers) and theoretical reconstruction via wavelet multiscale analysis (solid line) for $A=400\mu m^2$ and $\alpha=1$; (e) probability distribution $P(\delta\omega)$ of the resonances width $\delta\omega$ calculated for the fully chaotic resonator $\alpha=1$; (f) standard deviation $\sigma(\delta\omega)$ of the resonance widths versus $\alpha$.   
\label{exp1} 
}
\end{figure*}
The chaos-assisted energy buildup process observed when $\alpha\ge\alpha^*$ originates from the fundamental thermodynamic principle of equipartition, which can be highlighted using Eq. (\ref{tdcmt}). 
 By substituting $\tau_{k0}=\tau_0$ into the left hand of equations (\ref{tdcmt}), and assuming a dense 
 distribution of modes, with wavelength separation $\lambda_{k+1}-\lambda_k=d\lambda\ll\lambda $, we obtain:
\begin{align}
\label{eth}
\frac{\partial\mathcal{E}}{\partial\lambda}=\mathrm{const.}=\mathcal{E}_{0},
\end{align}  
which can be regarded as an equipartition theorem, with the energy equally 
distributed among all degrees of freedom ---i.e., the spectral wavelengths--- 
due to the strongly chaotic nature of the system. 
Equipartition is at the foundation of classical statistical mechanics, and forms the
 basis for thermodynamic ensembles and the observation of different phases of matter \cite{SMAST}. 
Applied to Photonics, we have the remarkable opportunity of exploiting this principle for enhancing the energy confinement properties of photonic structures. As already discussed above, this opportunity was confirmed 
by calculating the energy stored for a broadband source in the two limiting conditions of a fully chaotic geometry (Fig. \ref{eqp}a square markers) and a non-chaotic geometry (Fig. \ref{eqp}a circle markers), and finding a six-fold enhancement in the chaotic case. 

\begin{figure*}
\centering
\includegraphics[width=11cm]{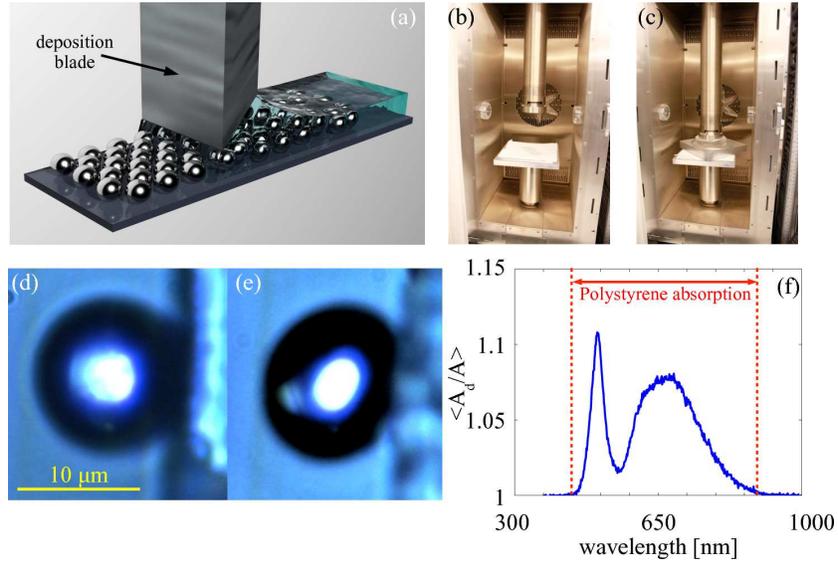}
\caption{Summary of the 3D experimental results with deformed microspheres: (a) sketch of mono-layer sample fabrication by convective self-assembly; (b)-(c) furnace for sample heating and deformation with mechanical pressure; (d)-(e) SEM micrograph of a microsphere in the original (d) and deformed (e) case; (f) average normalized absorption of the deformed microsphere $<A_d>=<A_d/A>$, measured for different wavelengths and normalized with respect to the undeformed case $A$.   
\label{exp2} 
}
\end{figure*}

\subsection*{2D experiments: Controlling the convergence towards a single modal lifetime} 

Energy equipartition and chaotic energy harvesting  both originate in the convergence of the modal lifetimes towards a single value $\tau_{k0}$. We therefore begin 
our experimental campaign by investigating the occurrence of this phenomenon in real structures. 
We designed and experimentally realized a series of planar 2D stadium-shaped resonators in planar photonic crystals (PhCs) fabricated in Silicon on Insulator (SOI). The substrate consists of a $220$nm thick silicon layer on a $2\mu$m thick insulator buried oxide. The patterns were written into ZEP resist on
a modified LEO/RAITH system with 2 nm step-size and etched with a 50:50 mixture of SF6 and CHF3 gases in a reactive ion etching machine. After stripping the residual resist, the sample was cleaved for end-fire coupling. Figures (\ref{exp1})a-c show a set of SEM micrographs and illustrate how the shape of the resonator evolves with a change of the deformation parameter $\alpha$. The shape starts as a regular square (Fig. \ref{exp1}a) for $\alpha$=0 and turns into a fully chaotic stadium-shaped resonator (Fig. \ref{exp1}c) for $\alpha$=1.5, with strong chaos already developed for $\alpha>\alpha^*=0.5$. The parameter $a=\sqrt{A/(4+\pi\alpha^2)}$ 
 guarantees a constant resonator area $A$ as $\alpha$ is varied (Fig. \ref{exp1}b). As the PhC lattice was designed to exhibit a bandgap around $1.5\mu$m with a bandwidth of $\approx 400$nm, any electromagnetic wave 
in this range is perfectly reflected at the PhC boundaries, and can only escape via the input/output waveguides of the structure or by scattering at imperfections. The system can therefore be described as a 2D resonator with measurable losses and the modal decay constants $\tau_k$ can 
 be extracted from the transmitted spectrum. We employed the multiscale analysis described in 
 \cite{falcoar:_lifet_statis_of_quant_chaos}, which provides an excellent reconstruction 
 technique even when the resonances overlap. The method fits the 
 spectrum via a sum of suitable wavelet functions, allowing to compute the wavelength position 
 $\lambda_0$ and the width $\delta\lambda$ of each mode. The resonance 
 widths $\delta\lambda$ are then inversely proportional to the modal decay rates 
 $1/\tau_k=\delta\omega=c\delta\lambda/n\lambda_0^2$  \cite{HausBook}. 
In order to collect statistically relevant data, we fabricated several samples 
 with different areas: $A=400\mu m^2$, $A=800\mu m^2$ and $A=1200\mu m^2$, and for 
 each area we considered five different degrees of chaos, expressed via the deformation parameters $\alpha=0$, $\alpha=0.25$, 
 $\alpha=0.5$, $\alpha=0.75$ and $\alpha=1$. To characterize the samples, we used polarized light in the wavelength range $1520$ nm to $1620$ nm. Figure \ref{exp1}d shows a portion of one of the measured spectra,
together with its reconstruction via multiscale analysis \cite{falcoar:_lifet_statis_of_quant_chaos}, which shows a perfect reproduction of the experimental results. All spectra (not shown here) have been reconstructed with the same level of accuracy. 
We are able to extract $\approx 2000$ resonances for each $\alpha$ 
, which allows us to extract statistically relevant trends. Figure \ref{exp1}e 
 displays the resonance linewidth probability distribution $P(\delta\omega)$ for $\alpha=1$ (stadium) and $\alpha=0$ (square). We note that for $\alpha=1$,
 the resonances are strongly converging towards a value of $\delta\omega$ $\approx 5\cdot 10^{-3}$ THz with only negligible 
 contributions arising from the short lived modes that are characterized by a larger $\delta\lambda$. Conversely, in the non chaotic regime, we observed the presence of many short lived resonances, indicated by the presence of data points up to, and even beyond 0.1 THz, as well as the wider probability distribution observed at low frequencies. In order to study the convergence of the lifetimes towards a single value, we group the data 
 for different $A$ and the same $\alpha$ together, and calculate the standard deviation $\sigma(\delta\omega)$ of the resonances width $\delta\omega$. 
Figure 
 \ref{exp1}f illustrates the results of this analysis. In perfect agreement with our theoretical 
 predictions, we observed a significant narrowing of the linewidth distribution above the threshold for chaos. (Fig. \ref{exp1}d). It is worthwhile highlighting that this convergence of the linewidth towards a single value does not 
 depend on $A$, but only on $\alpha$, which is a clear experimental demonstration that the 
 phenomenon relies entirely on the chaotic properties of the motion of light. 

\subsection*{3D experiments with deformed microspheres: chaos-enhanced broadband absorption of light}
A true comparison between model and experiment requires a direct measurement of the energy in the system.
We accomplished this final step in a three dimensional geometry, which incidentally also proves how the
physics of chaotic resonators is independent of the dimensionality of the problem. 
We employed polystyrene 
microspheres, and studied their absorption when their spherical shape gets deformed. Our sample consists of a low density monolayer of microspheres 
(Fig. \ref{exp2}a), which are being deformed by mechanical compression (Fig. \ref{exp2}b) in order 
to obtain an asymmetric shape such as that shown in Fig. \ref{sample}b. The monolayer fabrication was performed by 
employing convective assembly \cite{Malaquin2007, li2011}, where the microspheres are self-assembled on a substrate via a deposition blade (Fig. \ref{exp2}a).  The blade height was 
set at 12$\mu$m from the substrate (the sphere size is $\approx 10\mu$m) to ensure the formation of a single layer.  Following deposition,
a glass slide was placed on top of the microspheres for heating and applying mechanical pressure. 
The deformation was realized by heating the system slightly beyond the glass transition temperature 
$T_g$ of polystyrene ($T_g=94^\circ$ Celsius \cite{suh2004}), which softens the microspheres sufficiently to deform their shape (Fig. \ref{exp2}b-c). 
Compression and heating were performed with an Instron 5960 
dual column tabletop universal testing system. The pressure force was ramped to 500N over 2.5 mins. Figures \ref{exp2}d-e 
show optical micrographs of a microsphere before (d) and after (e) compression, 
highlighting the asymmetric deformation.\\
We evaluated the energy harvesting capacity of this system by performing absorption 
measurements on a single sphere in both deformed and undeformed conditions. In order to acquire a sufficiently large statistics from different input conditions, we placed our sample on a goniometric stage and, through a series of pump and probe measurements, we calculated the absorption of the sphere at various illumination angles in the range of $\pm 30$ degrees. During each measurement, we illuminated the sphere with a broadband source (with bandwidth $\approx 1\mu$m centered at $700$nm) and measured the absorption $A$ from the relation $A=1-T-R$, being $T$, $R$ transmittance and reflectance of the microsphere, 
 respectively. To properly collect the scattered light, transmission and reflection spectra were measured in the near field with a lens with a high numerical aperture. The data was acquired with an Ocean Optics QE65000-FL spectrometer. 
Figure \ref{exp2}f shows the average absorption $<A_d>=<A_d/A>$ of the deformed system, 
 being $A_d$ and $A$ the absorption measured in the deformed and undeformed case, respectively. For every angle we obtained an increased absorption due to deformation, with a variance within a few percent. This originates from the larger 
 electromagnetic energy stored by the microsphere in the deformed case, which led to a higher absorption 
 in the entire absorption window of polystyrene (Fig. \ref{exp2}f). Near the wavelength $\lambda=450$ nm, 
 in particular, the average absorption increases by $\approx 12\%$, while in the region 
 where the optical spectrum contains the maximum power ($\approx 600 nm$), the absorption 
 shows a broadband increase of a approximatively $6-8\%$. Outwith the absorption frequency 
 window of polystyrene (Fig. \ref{exp2} solid lines), as expected, the chaotic energy accumulation 
 does not induce any measurable variation of the absorption.

\subsection*{Discussion}
Our analytical, numerical and experimental results addressed the problem of light trapping in chaotically deformed resonators and demonstrate the possibility of exploiting chaos as a new avenue for energy 
harvesting. Our work began with a single numerical \emph{ab-initio} 
experiment, where we demonstrated broadband energy accumulation in an optical 
resonator once its shape was deformed to support the chaotic motion of light. We then developed an analytical approach based on a rigorous formulation of time dependent 
coupled mode theory, which highlights the principal phenomena of chaotic 
energy trapping. This approach allowed us to unveil the presence of the fundamental 
thermodynamic principle of energy equipartition. Equipartition, in our case, is manifested by the uniform distribution 
of energy among all degrees of freedom of the chaotic resonator, i.e. the cavity modes, and is evident from the convergence of the lifetime of the modes towards a single value.  A single lifetime means that each mode can store the same amount of energy, which overall leads to a significant energy buildup in the resonator. The process is equivalent to the Brownian motion of particles in a liquid, where each particle carries the same amount of energy.
The analogy goes even further. In a liquid, the particles always reach a uniform distribution regardless of the shape of the vessel they are contained in. In our optical analogue, the steady state of chaotic modes also does not depend on the particular 
realization of the resonator, but only on its 'macroscopic' geometry, whose only requirement 
is to support chaotic trajectories for the trapped light. A large series of 
parallel \emph{ab-initio} simulations perfectly confirmed our theoretical predictions.\\
We then conducted an experimental campaign to demonstrate both the main physical 
principles and the possible applications of this concept. In particular, we started by employing 
two-dimensional (2D) photonic crystals and realized stadium-shaped resonators with different degrees of chaoticity. 
Using pump and probe experiments, we demonstrated the convergence of the modal decay 
times towards a single value, which is at the basis of energy equipartition and chaotic energy harvesting. 
We then considered the realization of a practical three-dimensional (3D) geometry where we demonstrated enhanced light trapping through shape deformation \emph{directly}.  We employed polystyrene microspheres and altered their shape through heating and mechanical compression. We then measured the absorption 
of the system, and experimentally demonstrated a broadband increase across the entire
absorption window of the material by deforming the spheres form spherical to chaotic. The measured absorption increase was as high as $12\%$, which is quite 
remarkable considering the low absorption of polystyrene and the low refractive index of the material which only affords weak light trapping .\\
Besides the obvious implications at the fundamental level, where we demonstrate the existence of a fundamental principle of thermodynamics in the framework of Photonics, our results also have real-world practical implications. 
The cost of many semiconductor devices, e.g. LEDs and solar cells, is determined to a significant extent by the cost of the material. We show that the functionality of a given geometry, here exemplified by the energy that can be trapped in the system, can be enhanced up to six-fold by changing the shape alone, i.e. without increasing the amount of material and without increasing the material costs. Furthermore, a chaotic system is easier to fabricate as the tolerances are relaxed; the only requirement is that the resonator supports chaotic lightpaths, which is much easier to achieve in practise than manufacturing a resonator which is perfectly regular.

\begin{figure}
\centering
\includegraphics[width=7.5cm]{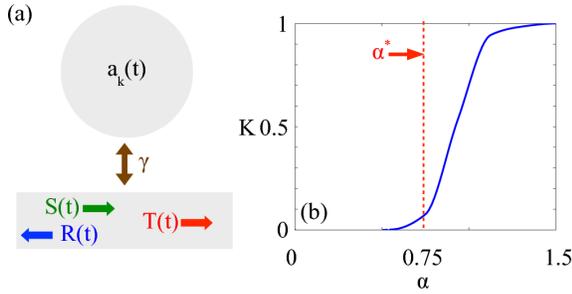}
\caption{(a) light-cavity interaction TDCMT scheme: $S(t)$ is the input source, $R(t)$, $T(t)$ are reflection and transmission signals, respectively, $a_k$ is the $k$-th mode in the resonator and $\gamma$ is the coupling coefficient between the resonator and the external environment; (b) normalized entropy $K$ versus chaotic degree $\alpha$ for the resonator defined by Eq. (\ref{curve}).    
\label{supp} 
}
\end{figure}

\subsection*{Methods}
\paragraph*{Ab-initio computations. ---} Numerical simulations were performed with our NANOCPP Maxwell solver (www.primalight.org). NANOCPP is a two/three-dimensional FDTD Maxwell code based on a parallel implementation of the Yee FDTD algorithm. The code employs uniaxial perfectly matched layers (UPML) to efficiently absorb outgoing waves and provides both hard and planewave sources [the latter via total field scattered field (TFSF) formulation]. NANOCPP efficiently scales over hundreds of thousands of processors, and is specifically designed for massively parallel electromagnetic computations. Our simulations consisted of a total of $20$ Million of single-cpu computational hours, with each simulation employing 4096-8192 computational cores on the KAUST Shaheen computational cluster and a resolution of at least $20$ points per internal wavelength. 

\paragraph*{Time dependent coupled mode equations. ---} 
Figure \ref{supp}a shows a TDCMT model of the light-resonator system. In this scheme, the environment is represented as a waveguide side coupled to the resonator. The input signal $S(t)$ interacts with the resonator through the coupling coefficient $\gamma$ and transfers energy to the modes of the cavity $a_k(t)=A_k e^{i\omega_k t-t/\tau_k}$ ($k\in[1,N]$), being $\omega_k$ the frequency, $\frac{1}{\tau_k}$ the lifetime, $A_k$ the amplitude and $\mathcal{E}_k=|a_k|^2$ the energy carried by the $k-$th mode. The total energy stored in the resonator is $\mathcal{E}=\sum_k\mathcal{E}_k$. Decay rates $\frac{1}{\tau_k}$ can be further decomposed as follows $\frac{1}{\tau_k}=\frac{1}{\tau_{k0}}+\frac{1}{\tau_e}$, with $\frac{1}{\tau_{k0}}$ the intrinsic decay constant of the mode and $\frac{1}{\tau_e}$ the decay constant due to coupling with the source. Cavity modes obey the following evolution equations  \cite{HausBook}:
\begin{align}
\label{tdcmt0}
&\frac{d a_k}{dt}=\bigg[i\omega_k-\bigg(\frac{1}{\tau_{k}}\bigg)\bigg]a_k+\gamma\cdot S(t), &k\in[1,...,n],
\end{align} 
with $\gamma=\sqrt{\frac{1}{\tau_e}}$ \cite{HausBook}, while reflection $R=\sum_k R_k$ and transmission $T=\sum_k T_k$ are given by the following expressions:
\begin{align}
\label{rt0}
&R_k=-\sqrt{\frac{1}{\tau_e}}a_k, &T_k=-\sqrt{\frac{1}{\tau_e}}+S,
\end{align}
with $R_k$ and $T_k$ the reflection and transmission of the $k-$th mode, respectively \cite{HausBook}.
In the presence of a single frequency excitation $S=e^{i\omega t}\Theta(t)$ switched on at $t=0$, with $\Theta(t)$ being the Heaviside function, Eqs. (\ref{tdcmt0}) are readily solved for each $a_k$ and read:
\begin{equation}
\label{ssol0}
a_k=\sqrt{\frac{1}{\tau_e}}\frac{e^{i\omega t}-e^{i\omega_kt-t/\tau_k}}{i(\omega-\omega_k)+\frac{1}{\tau_k}}.
\end{equation}
For a broadband source, $S=\int d\omega e^{i\omega t}\Theta(t)$, the total electromagnetic energy $\mathcal{E}=\sum_k|a_k|^2$ stored in the cavity is readily found to be:
\begin{align}
\label{tdcmt1}
\mathcal{H}=\int \frac{d \omega}{\tau_e}\sum_{k}\frac{1+e^{-\frac{2t}{\tau_k}}-2\cos[(\omega_k-\omega)t]e^{-\frac{t}{\tau_k}}}{(\omega_k-\omega)^2+\frac{1}{\tau_k^2}}.
\end{align}
Equation (\ref{tdcmt1}) can be further simplified as the integral yields significant contributions only for $\omega\approx\omega_k$, and we obtain:
\begin{equation}
\label{tdcmt2}
\mathcal{H}\tau_e\approx \sum_k\tau_k^2\big(1-e^{-\frac{t}{\tau_k}}\big)^2,
\end{equation}
which represents the first of Eq. (\ref{tdcmt}). The power $\mathcal{P}_k$ transferred into the $k-$th mode is conversely evaluated from the energy balance equation:
\begin{equation}
\frac{\partial |a_k|^2}{\partial t}=\mathcal{P}_k=|S|^2-|R_k|^2-|T_k|^2,
\end{equation}
that, in conjunction with  Eqs. (\ref{rt0})-(\ref{ssol0}), yields the second of Eqs. (\ref{tdcmt}).

\paragraph*{Characterization of chaos. ---}
We quantitatively describe the chaoticity of light motion by evaluating the relative area of the resonatorÕs phase space that encompasses chaos. This is achieved by first calculating the distribution of the Lyapunov exponent \cite{OTT} in phase space, and then performing a weighted summation by assigning '1' if the Lyapunov exponent is positive, and '0' otherwise. The resulting quantity $K$ can be regarded as a normalized version of the Kolmogorov-Sinai entropy \cite{Boffetta2002367,OTT}. In particular, when $K=0$, the resonator dynamics exhibits no chaos and the resulting motion is totally reversible, while for $K=1$ the phase space is totally chaotic and all input conditions lead to chaos. Values of $K$ between these two limiting conditions indicate a phase space partially chaotic, with $K$ measuring the relative area of the chaotic sea with respect to the reversible portion of the dynamics. Figure \ref{supp}b displays the behavior of $K$ for the billiard of Eq. \ref{curve}. For $\alpha$ lower than the threshold value $\alpha^*=0.75$, no chaos is observed in the structure, while for $\alpha>\alpha^*$ strong chaos is generated through the shape deformation. At $\alpha=1.5$, the structure is fully chaotic and the phase-space is totally dominated by a single chaotic sea.

\subsection*{Acknowledgments}
For computer time, this research used the resources of the 
Supercomputing Laboratory at King Abdullah University of Science \& Technology (KAUST) in 
Thuwal, Saudi Arabia.

\subsection*{Authors contributions}
A. Fratalocchi initiated the work and developed the theory behind chaotic energy harvesting. C. Liu and D. Molinari carried out numerical FDTD simulations and performed data analysis. A. Di Falco and T. F. Krauss realized the PhC samples and performed the experiments on the 2D geometries. B. S. Ooi and Y. Khan performed experiments on 3D deformed microspheres. All authors contributed to the manuscript preparation.

%

\end{document}